%% file: plb.tex
\documentclass[aps,prl,twocolumn,showpacs,superscriptaddress,groupedaddress]{revtex4}  
\usepackage{graphicx}  
\usepackage{dcolumn}   
\usepackage{bm}        
\usepackage{amssymb}   

\begin{document}

\hspace{5.2in} \mbox{FERMILAB-PUB-10-361-E}

\title{High mass exclusive diffractive dijet production in $\mathbf{p\bar{p}}$ collisions at $\mathbf{\sqrt{s}}$ = 1.96 TeV}
\input author_list.tex
\date {September 13, 2010}

\begin{abstract}
We present evidence for diffractive exclusive dijet production with an invariant dijet mass greater
than 100 GeV in data collected with the D0 experiment at the Fermilab Tevatron Collider.
A discriminant based on calorimeter information is used to
measure a significant number of events with little energy (typically less than 10 GeV) 
outside the dijet system, consistent with the diffractive exclusive dijet production
topology. The probability for these events to be
explained by other dijet production processes is $2 \times 10^{-5}$,
corresponding to a 4.1 standard deviation significance.
\end{abstract}

\maketitle

Hard diffraction was first observed 
about twenty years ago in the UA8 experiment at the CERN
$p\bar{p}$ collider SPS~\cite{bib:ua8}
and has been studied extensively in several experiments: H1 and ZEUS
at the DESY $ep$ Collider HERA~\cite{bib:h1zeus}
and D0 and CDF at the Fermilab Tevatron~\cite{bib:d0cdf}.
At hadron colliders, hard diffractive events are identified by 
the signature of a hard scatter in the presence
of a region devoid of any activity in the forward region of the
detector or by tagging beam hadrons
in the final state. Hard diffractive events can be described by
the exchange of a colorless object (Pomeron)~\cite{bib:diffreview}. 
Diffractively produced objects such as dijets, diphotons and $\chi_c$
charmonium can be 
observed in the
detector together with Pomeron remnants. 
A subset of hard diffractive
events in which both incoming hadrons remain intact
is defined in such a way that all the energy not
carried away by the outgoing beam particles is
used to produce the diffractive system~\cite{bib:original,bib:kmr}. 
This  mechanism is defined as hard exclusive diffractive production.
We search for this production mechanism in the
sample of dijet events with large dijet invariant mass,
corresponding to large values of the reduced center-of-mass
energy of the Pomeron's system.

Exclusive diffractive production (EDP) of a final state ${X}$,
$p\bar{p} \rightarrow p+X+\bar{p}$,
has been proposed as a future search channel for new 
physics, as well as for the Higgs boson, at the Large Hadron Collider (LHC)
\cite{bib:fp420}. In this process, the kinematic 
properties such as the mass of the object $X$ can be computed
with high precision by measuring only the momentum
loss of the outgoing protons in the final state. 
The CDF Collaboration reported the observation of exclusive diffractive 
events in the dijet, dielectron, diphoton and charmonium 
channels~\cite{bib:cdfobservation}.
These results support the existence of EDP,
but are restricted to low mass objects (typically less than 100~GeV),
while at the LHC, searches for new physics are expected to extend to higher masses.
In this article, we report evidence for exclusive diffractive dijet production with 
invariant masses greater than 100 GeV in data collected by the D0 experiment.

\begin{figure}
\includegraphics[width=0.32\textwidth]{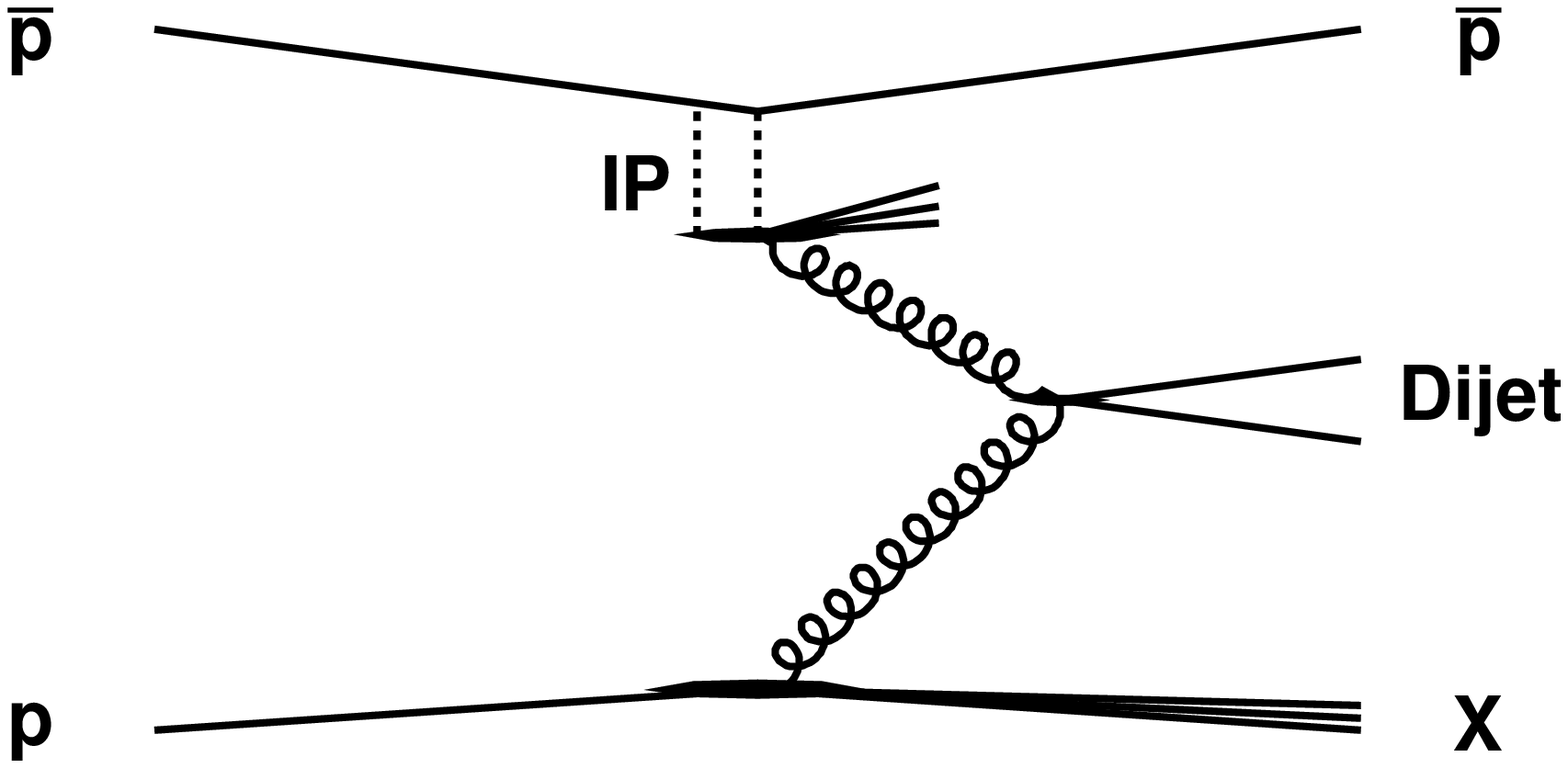} 
\\(a)\\ 
\includegraphics[width=0.32\textwidth]{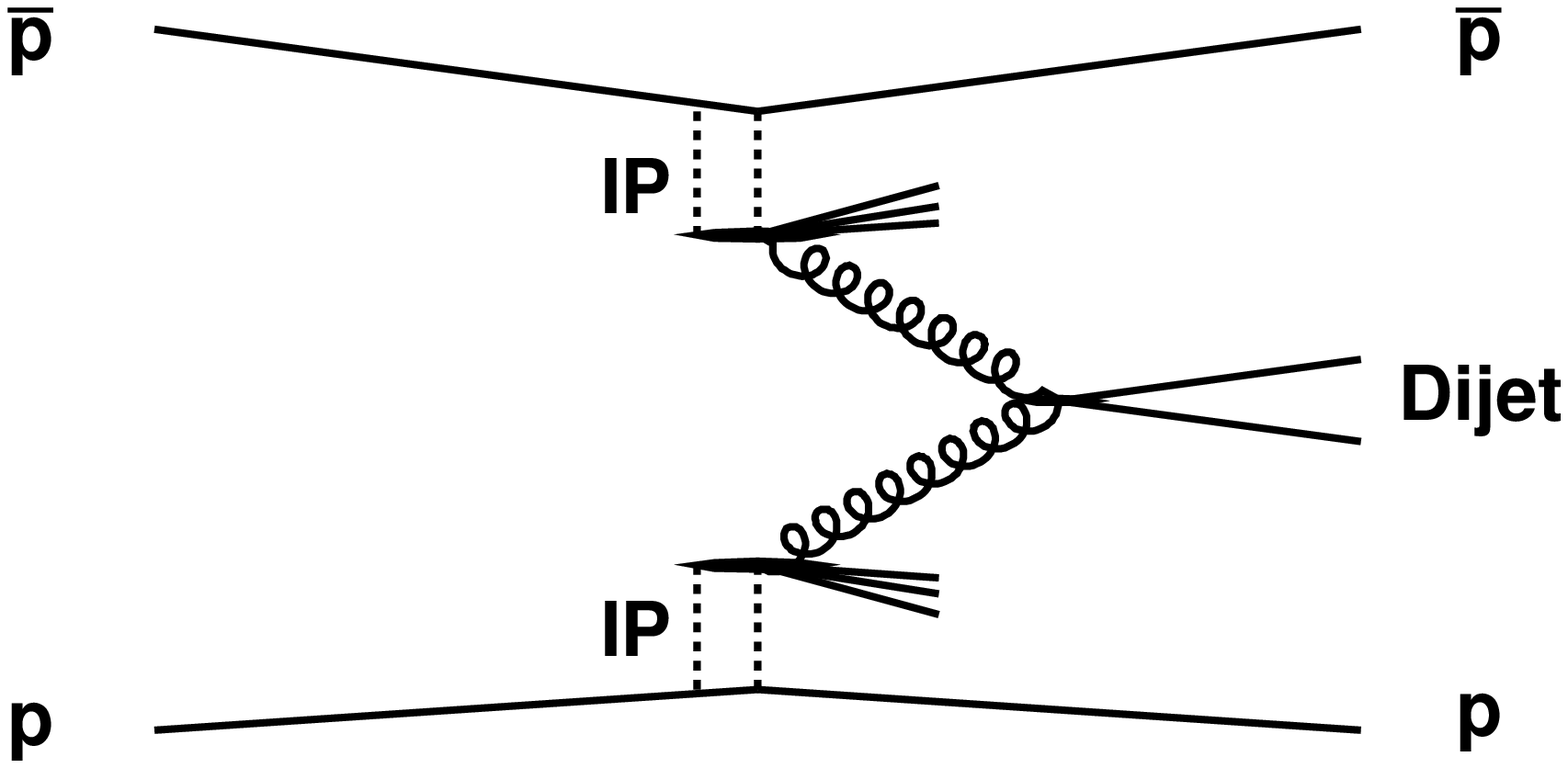}
\\(b)\\ 
\includegraphics[width=0.32\textwidth]{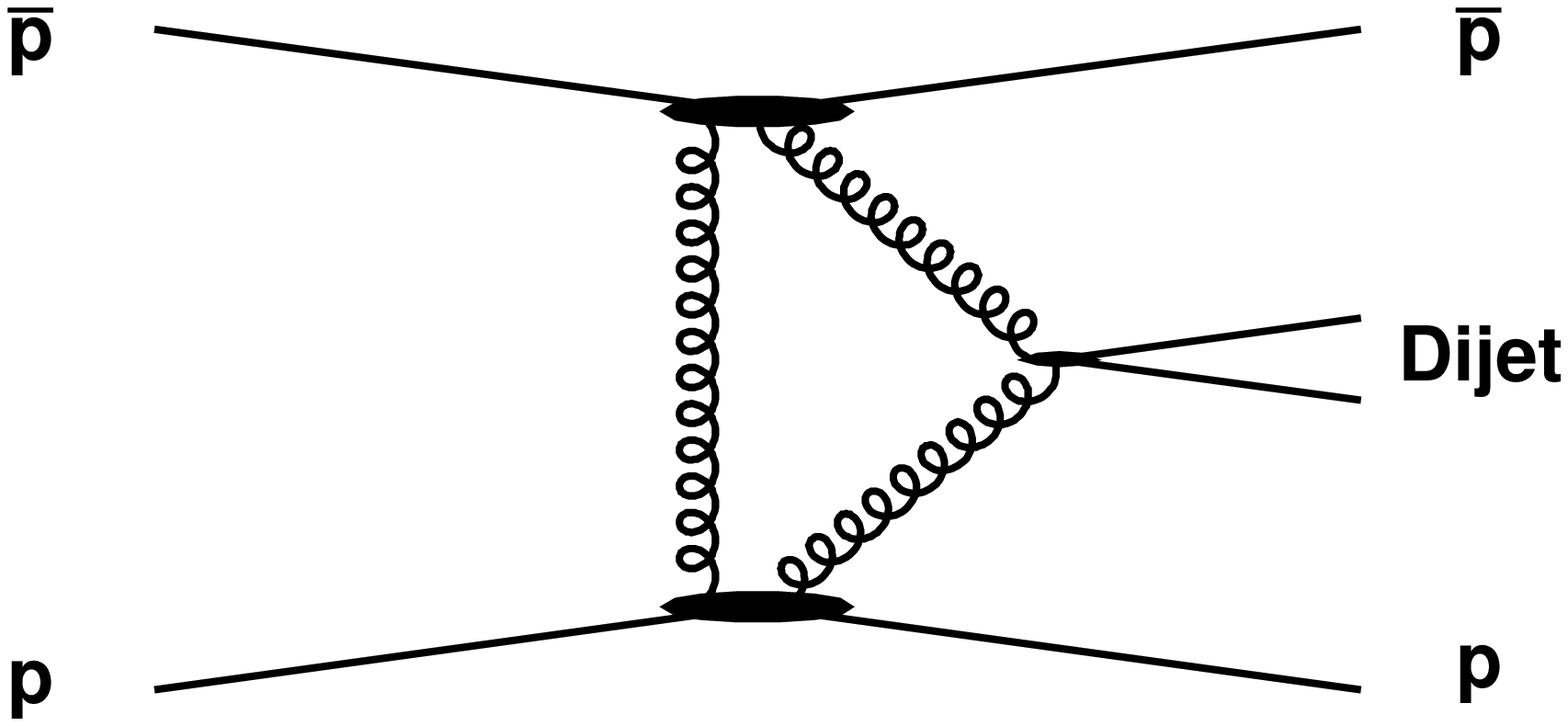}
\\(c)\\ 
\caption{Production of central dijet events in hard diffraction:
(a) single diffraction, in which only either the proton or the antiproton is diffracted
by a Pomeron ($I\!\!P$) exchange, while the other breaks up; 
(b) inclusive double Pomeron production, where
proton and antiproton remain intact, and additional QCD radiation
can be observed from Pomeron remnants; and 
(c) exclusive diffractive production where both protons remain intact and  only the 
dijet system is produced in the central region.
}
\label{fig:hardjj}
\end{figure}

We consider three different classes of
hard diffractive production in addition to non-diffractive production:
single diffractive (SD) dijet production (Fig.\ \ref{fig:hardjj} (a)), 
inclusive diffractive production through double Pomeron exchange (IDP) (Fig.\ \ref{fig:hardjj} (b)), and
exclusive diffractive dijet production (Fig.\ \ref{fig:hardjj} (c)).
In SD, one of the beam hadrons remains intact while the other breaks up.
In IDP, both beam hadrons are intact after the collision. 
The IDP and EDP processes with proton dissociations are expected to be 
suppressed by about a factor ten relative to the channel where the
beam hadrons remain intact~\cite{fpmc}.
The parton distributions of the
Pomeron are taken from recent H1 measurements~\cite{bib:hera} and
used to compute the diffractive jet production cross
section at the Tevatron. 
An additional multiplicative factor (gap survival probability)
~\cite{survival} of 0.1 is introduced to account for
soft production of particles from the underlying $p\bar{p}$ events
that populate the rapidity gaps~\cite{bib:diffreview}.

The background to EDP in the dijet mass region considered here 
originates from SD, IDP and non diffractive (NDF) events which have
either low multiplicity or small energy deposits in the forward calorimeters.
Due to the steeply falling nature of these distributions, these backgrounds are
expected to be small.
NDF background events
are simulated using the {\sc pythia} v6.202~\cite{pythia} Monte Carlo (MC) generator
with default settings and 
the diffractive (SD and IDP) backgrounds are determined using the
{\sc pomwig} v2.0~\cite{pomwig} and {\sc fpmc} v1.0~\cite{fpmc} generators respectively.
An EDP of dijet events at the lowest order of QCD~\cite{bib:kmr}
is simulated using {\sc fpmc} through the exchange of two gluons.

The data used in this analysis were collected with the D0 
detector in the period between August 2002 and April 2006 at the 
Tevatron Collider at a center-of-mass energy $\sqrt{s} = 1.96$ TeV.
The D0 detector is described 
in detail elsewhere \cite{d0det}. For this analysis, the most relevant components 
are the central and forward calorimeters used for jet reconstruction and 
the identification of a rapidity gap devoid of any energy (above noise) in the calorimeter,
respectively.
The D0 liquid argon and uranium calorimeter is divided in three parts
housed in independent cryostats covering the following
regions in pseudo-rapidity: 
$|\eta|<$1.1 (central calorimeter), and
$1.6<|\eta|<$ 4.2 (two forward calorimeters) where
$\eta= -\ln [\tan({\theta}/{2})]$ and
$\theta$ is the polar angle with respect to the beam axis.
Jets in EDP events are expected to be more central than in the other jet production
processes, 
therefore both jets are required to be central with a rapidity $|y|<0.8$,
where the rapidity is defined as $y=0.5 \ln (E+p_z)/(E-p_z)$ 
where $E$ is the jet energy and $p_z$
is the momentum component of the jet along the beam axis.
The forward region of the calorimeter is used to check for the presence
of a rapidity gap on each side of the dijet system.

\begin{figure}
\includegraphics[width=0.45\textwidth]{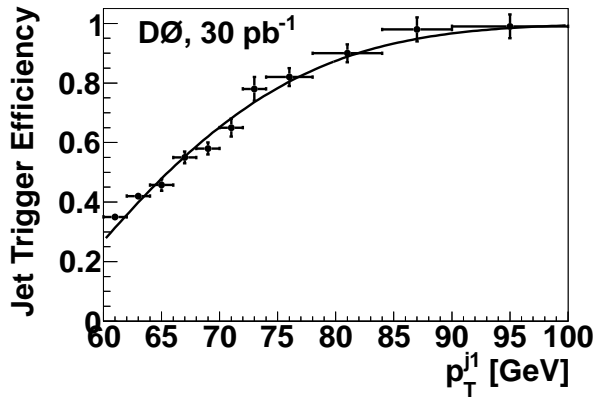}
\caption{Jet trigger efficiency as a function of
the leading jet $p_T$ ($p_T^{j1}$).
For events with $p_T^{j1} >$ 100 GeV, the efficiency is close to 100\% and no correction
is needed. 
\label{fig:trigger}}
\end{figure}

The instantaneous luminosity used in this analysis is required to
be in the range $[5-100]  \times 10^{30}$ cm$^{-2}$ s$^{-1}$, 
where the contribution from two or more $p\bar{p}$
interactions in a single event is in general much less than 20\%. This reduces the contamination
of multiple interactions in the same bunch crossing to the rapidity
gap selection.
Data were collected using a inclusive jet trigger requiring 
at least one jet in an event to be above a $p_T$ threshold of 45 GeV 
on the uncorrected energy, in order to
to select exclusive diffractive events in the region
of dijet invariant mass above 100 GeV.
Due to prescales imposed to avoid saturating the data acquisition system rate capabilities,
the equivalent integrated luminosity of the sample is about 30 pb$^{-1}$.
By comparing the highest-$p_T$ jet spectrum with data collected with a 
trigger with a lower $p_T$ threshold of 15~GeV, the trigger 
was found not to be
fully efficient for jet $p_T$ between 60~GeV and 100~GeV and 
the Monte Carlo events were reweighted with the trigger efficiency in this jet $p_T$ range.
The trigger efficiency as a function of jet $p_T$ is shown in Fig.\ \ref{fig:trigger}.

Jets are reconstructed using an iterative midpoint cone algorithm~\cite{d0jets}
with a cone size ${\cal R}=\sqrt{(\Delta \eta)^2+(\Delta \varphi)^2}=$  0.7,
where $\varphi$ is the azimuthal angle.
The highest-$p_T$ and second-highest $p_T$ jets 
are required to have $p_T$ greater 
than 60~GeV and 40~GeV, respectively, and only dijet events with an invariant mass
greater than 100 GeV are selected. 
To enhance the number of events without additional QCD 
radiation~\cite{bib:kmr},
the two jets are required to be
back-to-back in azimuthal angle $\varphi$, with a separation $\Delta \varphi > 3.1$.
A possible contribution of fake dijet events due to cosmic
rays is suppressed by the requirement that the missing
transverse momentum is less than 70\% of the leading jet transverse momentum.

\begin{figure}
\includegraphics[width=0.45\textwidth]{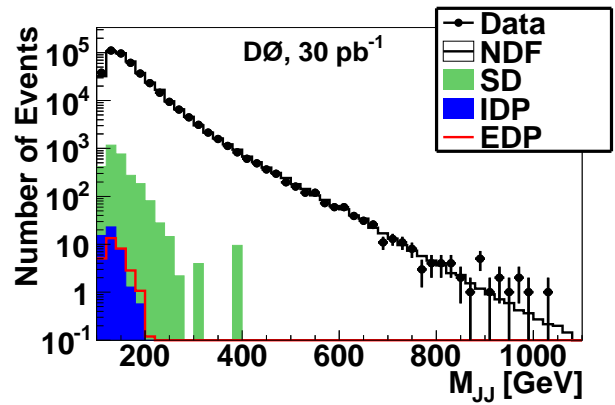}
\caption{Dijet invariant mass distribution for MC and data.
Good agreement between the MC simulation and data is found after applying jet
energy scale corrections and scale factors corresponding to the trigger efficiencies, the
luminosity profiles, and the MC normalization.
\label{fig:jetvar_all}}
\end{figure}

The MC events are required to satisfy the same selection criteria as the data.
They are processed through a GEANT-based~\cite{geant} simulation of the D0 detector
response and the same reconstruction code as data.
To simulate calorimeter noise and the effects of additional 
$p\bar{p}$ interactions, data events from random $p\bar{p}$ crossings are
overlaid on the MC events, using data from the same time
period as considered in the analysis.
The MC events are weighted to
obtain the same instantaneous luminosity profile as the data to have  
the same additional energy deposits in the forward region of the calorimeter 
as in data.
The sum of the number of NDF, SD and IDP events is normalized to
data and their fractions are determined by the theoretical
cross sections.
The EDP contribution is expected to be negligible at this stage.
Good agreement between the MC simulation and data is seen in
the dijet invariant mass distributions in Fig.\ \ref{fig:jetvar_all}.
By varying the requirement on the leading jet $p_T$, the uncertainty on the
normalization was estimated to be 5\%.

To discriminate between exclusive events and
background (NDF, SD and IDP) we exploit the large rapidity
gap that is expected between the central jets and the proton
and antiproton beams.
Two separate regions of pseudorapidity $\eta$ are defined in 
the calorimeter far from the two central jets. The very 
forward region ($3.0 < |\eta | < 4.2$) allows discrimination of diffractive events
(SD and IDP) from NDF events, which are accompanied by beam remnants 
in this region of the calorimeter.
The intermediate forward region ($2.0 < |\eta | < 3.0$) is used to identify 
EDP events, since they show larger rapidity gaps than SD and IDP. 
To prevent noise contamination in the calorimeter region under consideration, noisy cells in the forward
region of the calorimeters, which present an occupancy that differed by more
than five standard deviations from the average, are removed. 
The cell response in MC was also adjusted to data
by applying a MC-to-data correction factor for each cell.
This correction factor
was obtained using data collected requiring either
minimal activity in the D0 luminosity counters or
the presence of low $p_T$ jets.
After performing these corrections, the calorimeter cell information was used
to form the following 
variable in order to discriminate between the different 
classes of events

\begin{equation}
\label{eq:vare}
\Delta =  
\frac{1}{2} \exp(-\sum_{2 < |\eta |< 3} E_{T}) + 
\frac{1}{2} \exp(-\sum_{3 < |\eta |< 4.2} E_{T}).
\end{equation}

\begin{figure}
\includegraphics[width=0.45\textwidth]{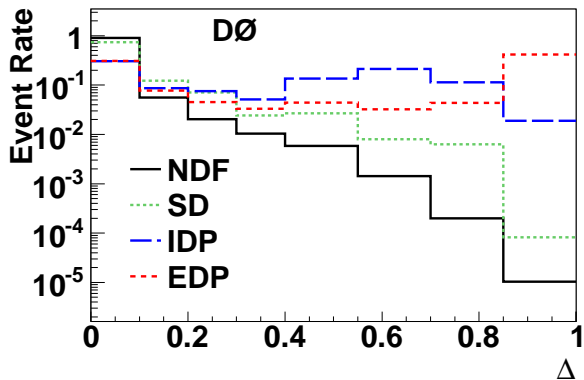}
\caption{Distribution of $\Delta$ normalized to unity for all MC samples. EDP
peaks at $\Delta > 0.85$.
The EDP contribution at low $\Delta$ values is due to pile-up 
events, where a second proton-antiproton inelastic scattering
occurs in the same bunch crossing.
\label{fig:variables}}
\end{figure}

Here $E_T$ is the transverse energy in a given  cell, and the
sum is performed over all cells in the indicated rapidity range.
Figure \ref{fig:variables} displays the $\Delta$ distribution normalized to unity 
for all MC samples. Also shown is the expected distribution
from EDP events, showing a characteristic peak at $\Delta>0.85$,
corresponding to energy deposits in the forward calorimeter
which are typically smaller than 10 GeV.

\begin{figure}
\begin{center}
\includegraphics[width=0.45\textwidth]{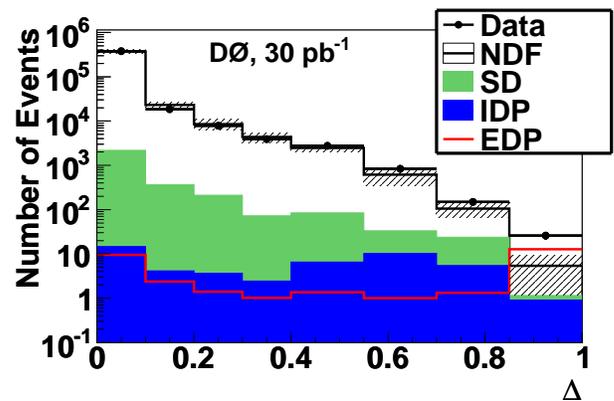}
\\(a)\\
\includegraphics[width=0.45\textwidth]{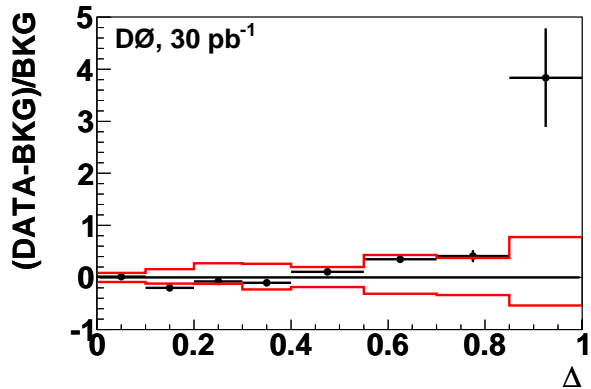}
\\(b)\\
\caption{ 
(a) Distribution of $\Delta$ for data and contributions from NDF, SD and IDP. 
The hatched band indicates the total uncertainty on the background. 
(b) Normalized difference between data and NDF, SD and IDP processes. 
The red solid lines are $\pm$ 1 standard deviation systematic uncertainties on the background.
\label{fig:vare_all}}
\end{center}
\end{figure}

Systematic uncertainties are assessed on the MC background 
prediction of the differential distribution for $\Delta$.
The leading systematic uncertainty is due to the 
calorimeter cells calibration factors.
They are varied simultaneously by three standard deviations from their central
value leading to a change of 25\% of the background for $\Delta >$ 0.85.
The effect of the jet energy scale uncertainties modifies the background by 12\%. 
The jet energy resolution in simulation has been varied to match the data, yielding
a small change of the normalization of $0.5\%$  which is assigned as an uncertainty.
To estimate the uncertainties of the
trigger efficiency correction and the instantaneous luminosity reweighting,
the analysis was repeated using a 15 GeV jet $p_T$ trigger threshold
resulting in a 3\% systematic shift. 
An additional systematic uncertainty 
due to the MC to data normalization 
is estimated to be 5\%. 
An uncertainty of 50\% on the SD and IDP MC cross sections accounts for
the uncertainty on the partonic structure of the Pomeron and survival
probability gap factor.
The non-diffractive parton distribution function uncertainties
were considered and found to be negligible with respect to the
other uncertainties.
The total background prediction is 5.4 $^{+4.2} _{-2.9}$ events and
26 signal candidate events are observed in data.

\begin{figure}
\begin{center}
\includegraphics[width=0.45\textwidth]{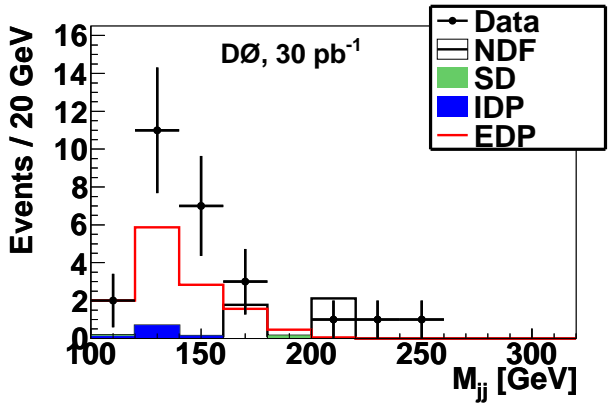}
\caption{Dijet invariant mass distribution for 
stacked MC (NDF, SD and IDP) and data after
applying the requirement on $\Delta$ $\geq$ 0.85.
The EDP distribution is shown without adding the stacked background.
\label{fig:jetvar_signal}}
\end{center}
\end{figure}

\begin{figure*}
\begin{center}
\includegraphics[width=0.45\textwidth]{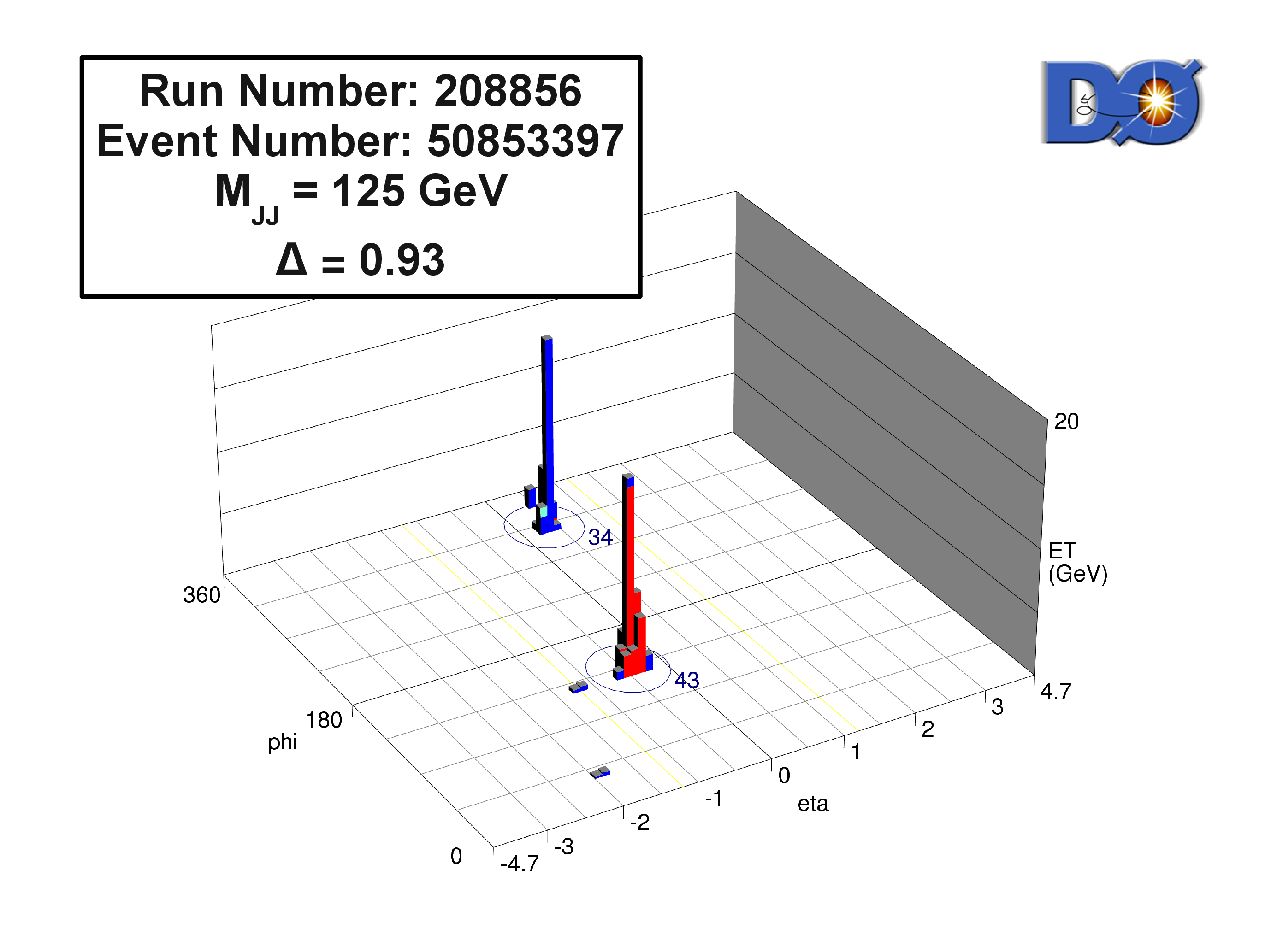} 
\includegraphics[width=0.45\textwidth]{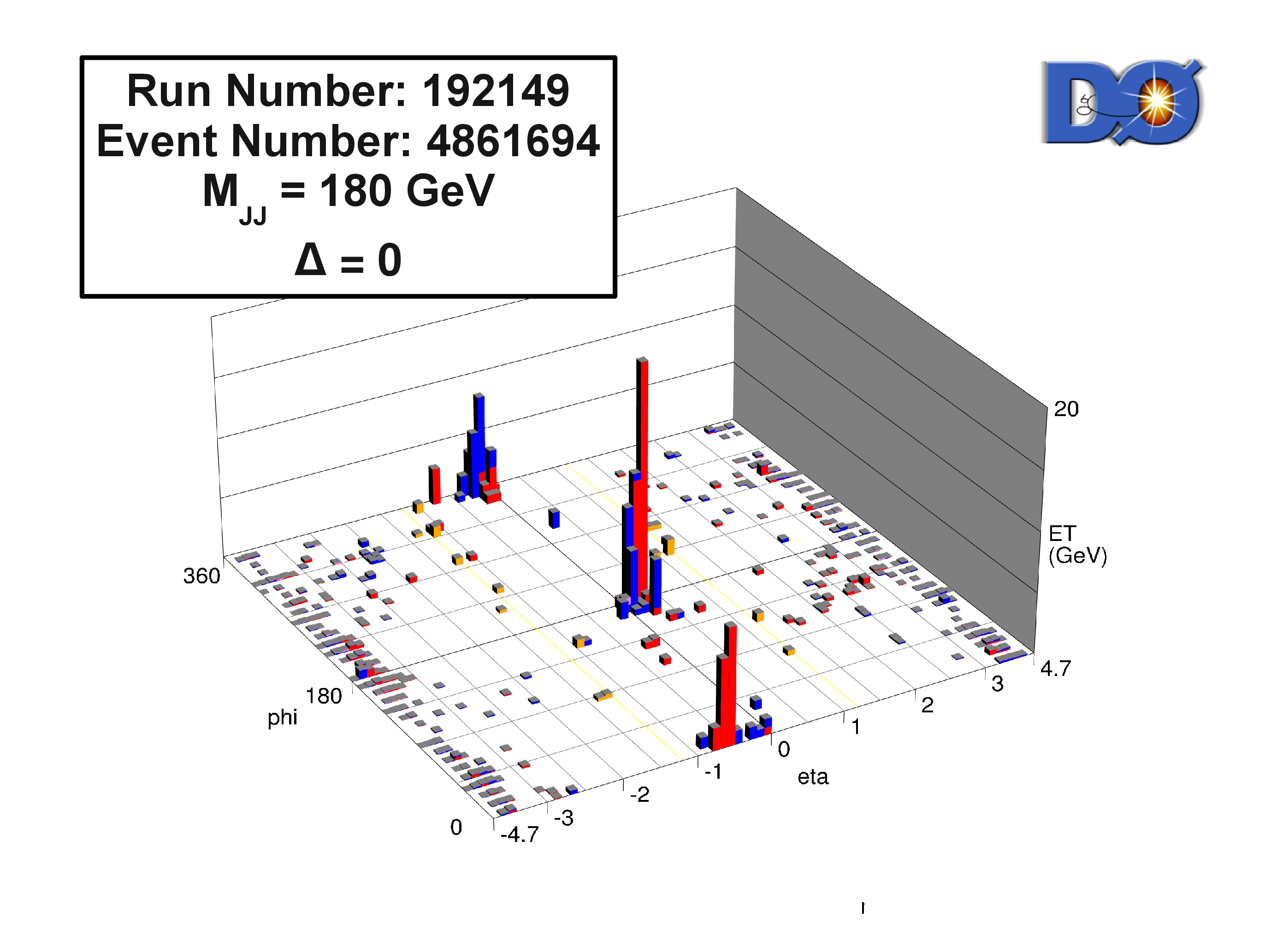} \\
(a) \hspace{10cm} (b) 
\caption{Event displays showing $E_T$ in the
$\eta-\varphi$ plane:
(a) Exclusive diffractive event candidate: No energy
deposition is present in the forward regions, only two central
jets are observed in the detector.
(b) Background event: In addition to
the two jets present in the detector, energy deposition is present 
in the forward regions.
The different colors correspond to energy deposits in different
layers of the calorimeter.
\label{fig:evtdisplay}}
\end{center}
\end{figure*}

Figure \ref{fig:vare_all} shows the comparison of the $\Delta$ distributions 
in data and MC (NDF, SD and IDP) normalized to their leading order cross
sections.
Good agreement is observed between data and MC except at high values of
$\Delta$ where EDP dominates.
The significance of the excess with respect to the NDF, SD and IDP backgrounds
is determined using a modified frequentist method~\cite{collie}.
It is obtained via fits of the signal$+$background and background-only hypotheses 
to pseudo-data samples containing only background. The effect of systematic
uncertainties is constrainted by maximizing a likelihood function 
for background and signal$+$background hypotheses over all systematic
uncertainties. Pseudo experiments used to determine the significance
of the EDP signal include variations over each
systematic uncertainty.
The observed significance corresponds to the fraction of outcomes 
that yield an EDP cross section at least as large as that measured in data.
Seven bins are used as input for the significance
calculation: six bins for $\Delta$ between 0.1 and 0.85, where the predominant region 
used in the MC normalization is removed, and
the $\Delta$ $\geq$ 0.85 bin. The probability for the
observed excess to be explained by  an upward fluctuation
of the background
is $2 \times 10^{-5}$, corresponding to an excess of $4.1$ standard deviations.
Table~\ref{tab:final} gives the observed number of events compared to
background and EDP expectations.
Figure \ref{fig:jetvar_signal} displays
the dijet invariant
mass distribution for $\Delta >$ 0.85. 
To illustrate the differences between the diffractive dijet exclusive events 
with $\Delta >$ 0.85,
where the calorimeter has little energy deposition outside
the central region, 
and the non-diffractive events, two
event displays are shown in Fig.\ \ref{fig:evtdisplay}.
The excess in data can contain events where the proton is dissociated into low-mass
states that escape detection. The contribution of such events is estimated to be up
to 10\% of the EDP cross section~\cite{fpmc}.

\begin{table*}
\begin{center}
\begin{tabular}[t]
{l || c | c | c | c | c | c }
\hline
Sample & NDF & IDP & SD & EDP & BKG & DATA \\
\hline\hline
All $\Delta$ & 409527 & 48.3 & 2930 & 30.9 & 412505 & 412505 \\
\hline
$\Delta$ $\geq$ 0.85 & 4.2 $^{+4.0} _{-2.9}$ & 0.9 $^{+0.4} _{-0.5}$  & 0.2 $^{+0.1} _{-0.1}$ & 12.9 $^{+1.0} _{-1.2}$ & 5.4 $^{+4.2} _{-2.9}$ & 26\\
\hline
\end{tabular}
\end{center}
\caption
{Number of predicted events for each MC sample for all $\Delta$ and for
$\Delta$ $\geq$ 0.85.
The total uncertainties are quoted.
\label{tab:final}}
\end{table*}

To summarize, we have presented evidence at the $4.1$ standard deviation level for 
events consistent with the exclusive dijet production event topology
in $p\bar{p}$ collisions 
at a center-of-mass energy $\sqrt{s} = 1.96$ TeV at 
high dijet invariant mass ($M_{JJ}> 100$ GeV).
These are the highest mass states studied for exclusive
production in hadron colliders.
Such event signatures are expected to play an important role in future studies at the Tevatron and
LHC.

\input acknowledgement.tex   

\end{document}

%% file: author_list.tex
\affiliation{Universidad de Buenos Aires, Buenos Aires, Argentina}
\affiliation{LAFEX, Centro Brasileiro de Pesquisas F{\'\i}sicas, Rio de Janeiro, Brazil}
\affiliation{Universidade do Estado do Rio de Janeiro, Rio de Janeiro, Brazil}
\affiliation{Universidade Federal do ABC, Santo Andr\'e, Brazil}
\affiliation{Instituto de F\'{\i}sica Te\'orica, Universidade Estadual Paulista, S\~ao Paulo, Brazil}
\affiliation{Simon Fraser University, Vancouver, British Columbia, and York University, Toronto, Ontario, Canada}
\affiliation{University of Science and Technology of China, Hefei, People's Republic of China}
\affiliation{Universidad de los Andes, Bogot\'{a}, Colombia}
\affiliation{Charles University, Faculty of Mathematics and Physics, Center for Particle Physics, Prague, Czech Republic}
\affiliation{Czech Technical University in Prague, Prague, Czech Republic}
\affiliation{Center for Particle Physics, Institute of Physics, Academy of Sciences of the Czech Republic, Prague, Czech Republic}
\affiliation{Universidad San Francisco de Quito, Quito, Ecuador}
\affiliation{LPC, Universit\'e Blaise Pascal, CNRS/IN2P3, Clermont, France}
\affiliation{LPSC, Universit\'e Joseph Fourier Grenoble 1, CNRS/IN2P3, Institut National Polytechnique de Grenoble, Grenoble, France}
\affiliation{CPPM, Aix-Marseille Universit\'e, CNRS/IN2P3, Marseille, France}
\affiliation{LAL, Universit\'e Paris-Sud, CNRS/IN2P3, Orsay, France}
\affiliation{LPNHE, Universit\'es Paris VI and VII, CNRS/IN2P3, Paris, France}
\affiliation{CEA, Irfu, SPP, Saclay, France}
\affiliation{IPHC, Universit\'e de Strasbourg, CNRS/IN2P3, Strasbourg, France}
\affiliation{IPNL, Universit\'e Lyon 1, CNRS/IN2P3, Villeurbanne, France and Universit\'e de Lyon, Lyon, France}
\affiliation{III. Physikalisches Institut A, RWTH Aachen University, Aachen, Germany}
\affiliation{Physikalisches Institut, Universit{\"a}t Freiburg, Freiburg, Germany}
\affiliation{II. Physikalisches Institut, Georg-August-Universit{\"a}t G\"ottingen, G\"ottingen, Germany}
\affiliation{Institut f{\"u}r Physik, Universit{\"a}t Mainz, Mainz, Germany}
\affiliation{Ludwig-Maximilians-Universit{\"a}t M{\"u}nchen, M{\"u}nchen, Germany}
\affiliation{Fachbereich Physik, Bergische  Universit{\"a}t Wuppertal, Wuppertal, Germany}
\affiliation{Panjab University, Chandigarh, India}
\affiliation{Delhi University, Delhi, India}
\affiliation{Tata Institute of Fundamental Research, Mumbai, India}
\affiliation{University College Dublin, Dublin, Ireland}
\affiliation{Korea Detector Laboratory, Korea University, Seoul, Korea}
\affiliation{CINVESTAV, Mexico City, Mexico}
\affiliation{FOM-Institute NIKHEF and University of Amsterdam/NIKHEF, Amsterdam, The Netherlands}
\affiliation{Radboud University Nijmegen/NIKHEF, Nijmegen, The Netherlands}
\affiliation{Joint Institute for Nuclear Research, Dubna, Russia}
\affiliation{Institute for Theoretical and Experimental Physics, Moscow, Russia}
\affiliation{Moscow State University, Moscow, Russia}
\affiliation{Institute for High Energy Physics, Protvino, Russia}
\affiliation{Petersburg Nuclear Physics Institute, St. Petersburg, Russia}
\affiliation{Stockholm University, Stockholm and Uppsala University, Uppsala, Sweden }
\affiliation{Lancaster University, Lancaster LA1 4YB, United Kingdom}
\affiliation{Imperial College London, London SW7 2AZ, United Kingdom}
\affiliation{The University of Manchester, Manchester M13 9PL, United Kingdom}
\affiliation{University of Arizona, Tucson, Arizona 85721, USA}
\affiliation{University of California Riverside, Riverside, California 92521, USA}
\affiliation{Florida State University, Tallahassee, Florida 32306, USA}
\affiliation{Fermi National Accelerator Laboratory, Batavia, Illinois 60510, USA}
\affiliation{University of Illinois at Chicago, Chicago, Illinois 60607, USA}
\affiliation{Northern Illinois University, DeKalb, Illinois 60115, USA}
\affiliation{Northwestern University, Evanston, Illinois 60208, USA}
\affiliation{Indiana University, Bloomington, Indiana 47405, USA}
\affiliation{Purdue University Calumet, Hammond, Indiana 46323, USA}
\affiliation{University of Notre Dame, Notre Dame, Indiana 46556, USA}
\affiliation{Iowa State University, Ames, Iowa 50011, USA}
\affiliation{University of Kansas, Lawrence, Kansas 66045, USA}
\affiliation{Kansas State University, Manhattan, Kansas 66506, USA}
\affiliation{Louisiana Tech University, Ruston, Louisiana 71272, USA}
\affiliation{University of Maryland, College Park, Maryland 20742, USA}
\affiliation{Boston University, Boston, Massachusetts 02215, USA}
\affiliation{Northeastern University, Boston, Massachusetts 02115, USA}
\affiliation{University of Michigan, Ann Arbor, Michigan 48109, USA}
\affiliation{Michigan State University, East Lansing, Michigan 48824, USA}
\affiliation{University of Mississippi, University, Mississippi 38677, USA}
\affiliation{University of Nebraska, Lincoln, Nebraska 68588, USA}
\affiliation{Rutgers University, Piscataway, New Jersey 08855, USA}
\affiliation{Princeton University, Princeton, New Jersey 08544, USA}
\affiliation{State University of New York, Buffalo, New York 14260, USA}
\affiliation{Columbia University, New York, New York 10027, USA}
\affiliation{University of Rochester, Rochester, New York 14627, USA}
\affiliation{State University of New York, Stony Brook, New York 11794, USA}
\affiliation{Brookhaven National Laboratory, Upton, New York 11973, USA}
\affiliation{Langston University, Langston, Oklahoma 73050, USA}
\affiliation{University of Oklahoma, Norman, Oklahoma 73019, USA}
\affiliation{Oklahoma State University, Stillwater, Oklahoma 74078, USA}
\affiliation{Brown University, Providence, Rhode Island 02912, USA}
\affiliation{University of Texas, Arlington, Texas 76019, USA}
\affiliation{Southern Methodist University, Dallas, Texas 75275, USA}
\affiliation{Rice University, Houston, Texas 77005, USA}
\affiliation{University of Virginia, Charlottesville, Virginia 22901, USA}
\affiliation{University of Washington, Seattle, Washington 98195, USA}
\author{V.M.~Abazov} \affiliation{Joint Institute for Nuclear Research, Dubna, Russia}
\author{B.~Abbott} \affiliation{University of Oklahoma, Norman, Oklahoma 73019, USA}
\author{M.~Abolins} \affiliation{Michigan State University, East Lansing, Michigan 48824, USA}
\author{B.S.~Acharya} \affiliation{Tata Institute of Fundamental Research, Mumbai, India}
\author{M.~Adams} \affiliation{University of Illinois at Chicago, Chicago, Illinois 60607, USA}
\author{T.~Adams} \affiliation{Florida State University, Tallahassee, Florida 32306, USA}
\author{G.D.~Alexeev} \affiliation{Joint Institute for Nuclear Research, Dubna, Russia}
\author{G.~Alkhazov} \affiliation{Petersburg Nuclear Physics Institute, St. Petersburg, Russia}
\author{A.~Alton$^{a}$} \affiliation{University of Michigan, Ann Arbor, Michigan 48109, USA}
\author{G.~Alverson} \affiliation{Northeastern University, Boston, Massachusetts 02115, USA}
% ...
%
%
\collaboration{The D0 Collaboration\footnote{with visitors from
$^{a}$Augustana College, Sioux Falls, SD, USA,
$^{b}$The University of Liverpool, Liverpool, UK,
$^{c}$SLAC, Menlo Park, CA, USA,
$^{d}$ICREA/IFAE, Barcelona, Spain,
$^{e}$Centro de Investigacion en Computacion - IPN, Mexico City, Mexico,
$^{f}$ECFM, Universidad Autonoma de Sinaloa, Culiac\'an, Mexico,
and 
$^{g}$Universit{\"a}t Bern, Bern, Switzerland.%
}} \noaffiliation
\vskip 0.25cm

%% file: acknowledgement.tex
%
We thank the staffs at Fermilab and collaborating institutions,
and acknowledge support from the
DOE and NSF (USA);
CEA and CNRS/IN2P3 (France);
FASI, Rosatom and RFBR (Russia);
CNPq, FAPERJ, FAPESP and FUNDUNESP (Brazil);
DAE and DST (India);
Colciencias (Colombia);
CONACyT (Mexico);
KRF and KOSEF (Korea);
CONICET and UBACyT (Argentina);
FOM (The Netherlands);
STFC and the Royal Society (United Kingdom);
MSMT and GACR (Czech Republic);
CRC Program and NSERC (Canada);
BMBF and DFG (Germany);
SFI (Ireland);
The Swedish Research Council (Sweden);
and
CAS and CNSF (China).